# A REVIEW ON COOPERATIVE DIVERSITY TECHNIQUES BYPASSING CHANNEL ESTIMATION

*Sylvia Ong Ai Ling, Hushairi Zen, Al-Khalid B Hj Othman, Mahmood Adnan and Olalekan Bello
Faculty of Engineering, University Malaysia Sarawak, Kota Samarahan, Kuching, Sarawak, Malaysia

## ABSTRACT

Wireless communication technology has seen a remarkably fast evolution due to its capability to provide a quality, reliable and high-speed data transmission amongst the users. However, transmission of information in wireless channels is primarily impaired by deleterious multipath fading, which affects the quality and reliability of the system. In order to overcome the detrimental effects of fading, Multiple-Input Multiple-Output (MIMO) technology is an attractive scheme that employs multiple transceiver antennas to carry the data over the same frequency band over a variety of signal paths. This technology has shown great solutions due to its ability to provide better spectral efficiency, capacity, throughput and robustness of the data transmission. But in practice, it is impractical to install multiple antennas on small-sized devices. Hence, to overcome the limitations of MIMO gain in the future wireless networks, cooperative diversity has recently draw in attention due to its ability to circumvent the difficulties of implementing actual antenna arrays in Multiple-Input and Multiple-Output (MIMO). By exploiting the broadcast feature of the wireless medium, cooperation among multiple nearby nodes is formed for data transmission. At the receiver, the signals are either coherently or differentially detected. Coherent detection requires exact channel estimation, which is difficult to apply in a time-varying channel. Hence, when the nodes are mobile, or when the channel is inaccurately estimated, the differential detection techniques that omit channel estimation become an alternative as compared to coherent detection. This article presents a review of the differential transmission techniques for cooperative diversity networks. The article begins with the development of the differential detection techniques. Then, the concept of double-differential technique with the presence of carrier offset is addressed. The review of these studies is presented so as to provide directions for future developments.

## INTRODUCTION

Over the past few decades, wireless communication has gained much popularity and became one of the most vibrant domains in the communications field, due to its rapid evolution to provide untethered connectivity. In order to provide high spectral efficiency and reliable high-speed communication links, the transmission system needs to be designed in such an efficacious manner that the available space and time is utilized as precisely as possible. This poses a challenging task as the wireless channel is susceptible to small-scale effect of multipath fading and large-scale fading which results in random variation of channel quality (Skylar, 1988). Multipath fading takes place when the signals propagate along different paths from the source to the destination. Such phenomenon transpires as a result of atmospheric scattering, diffraction, and reflection from obstructing objects (Proakis, 2001). The replica signals received at the receiver experiences different levels of attenuation and time delays that deteriorates the received signals.

Hence, diversity is an effective mechanism to combat the deleterious effects of the multipath fading using multiple independent communication channels (Proakis, 2001). Among various diversity-based technologies, MIMO system is often desirable due to its capability for spatial multiplexing reuse and multiple broadcast features (Nosratinia*et al*., 2004). Yet, it is impractical to employ multiple antennas in pocket-size transceiver sowing to the size, cost and hardware constraints, for instance, in wireless cellular networks and wireless sensor networks (Sendonaris *et al*., 2003). In such cases, an effective yet reliable cooperative diversity appears to be one of the promising techniques to meet the challenges to mitigate fading with low power consumption at the source. By means of applying the wireless medium broadcast characteristic, wherein some information is "overheard" within the nearby nodes and the capability to attain diversity through independent channels – cooperation amongst multiple neighbouring nodes can be realized in the virtual antenna sharing environment by employing

Time Division Multiple Access (TDMA). During the first time slot, the source broadcasts signals directly following independent wireless paths towards both the relay(s) and receiver(s). In the second time slot, the relay(s) process(es) the source signals using relaying protocols and retransmit it to the intended destination. At the destination, all the relayed signals and directly transmitted signals are detected and combined using various detection and diversity combining techniques to improve the performance of the faded received signals (Nosratinia *et al*., 2004).

While the cooperative diversity can significantly enhance the wireless communications networks' performance, existing studies have primarily concentrated on the coherent detection. In coherent schemes, it is required that Channel State Information (CSI) and synchronization of each link are perfectly known at the relay(s) and destination (Laneman and Wornell, 2003; Laneman *et al*., 2004; Farhadi and Beaulieu, 2008). Perfect channel knowledge is likely to be obtained from the conventional practical channel estimation schemes as highlighted in (Proakis, 2001) by implementing pilot symbols, which are used as reference signals for initial channel estimation and for initial synchronization. However, the extra pilot symbols transmission results in an undesirable wastage of the transmission power as well as the channel bandwidth and reduction of the network throughput (Annavajjala *et al*., 2005). Additionally, in actual practice, the wireless channel changes rapidly owing to the mobility of the nodes. In these circumstances, the amplitude and phase vary significantly all over the period of transmission, whereby the precise channel estimation becomes impossible. Hence, the channel estimation increases the system's complexity and error, especially in multiple nodes when the fast fading becomes apparent (Zhu *et al*., 2007; Tian *et al*., 2011).

Alternatively, in order to bypass the challenging and highly-complex channel estimation for wireless communication networks, the interest herein lies in the development of the differential cooperative transmission techniques (Tarokh and Jafarkhani, 2000; Tarasak *et al*., 2005; Himsoon *et al*., 2008; Zhao and Li, 2007). Since these techniques do not require any knowledge of the channel, the transmission bandwidth and power can be fully utilized without applying the pilot symbols transmission. Several differential cooperative transmission techniques have been proposed to omit the channel estimation requirement, both at the relay and destination. For differential schemes, Differential Phase-Shift Keying (DPSK) is a superior solution that can be demodulated at the destination without the requirement of channel estimation. Although channel estimation and pilot symbols are omitted, the differential encoding technique however incurred a 3 dB performance loss as compared to the coherent detection (Tarokh and Jafarkhani, 2000) in a non-frequency selective channel. However, in real practice, the channel is exposed to various impairments that can never be perfectly estimated at the destination which degrades the performance of the transmission links. On the other hand, it was revealed that multiple-symbol differential based detection is able to reduce the performance gap between the differentially encoded detection and the coherent detection schemes in (Divsalar and Simon, 1990). A Multiple-Symbol Differential Detector (MSDD) makes a joint-detection and decision for a block of $N$ consecutive received Pulse Shift Keying (PSK)[1] symbols based on $N+1$ samples. Lampe *et al*. (2005) further proposed the Sphere Decoding (SD) for multiple differential detection that outperforms the MSDD and yet, keeps the detection complexity level low.

However, existing studies have revealed that most proposed receivers primarily assumed detection without considering Carrier Frequency Offset (CFO). The presence of CFO is caused by the relative motion between the source and destination (i.e. Doppler shift) and the mismatch between the source and destination oscillators. In the presence of CFO, the fading channel varies and leads to performance degradation of the differential detection. To date, this is a vital practical problem for cooperative diversity system which has not yet been given much attention. Hence Double-Differential (DD) modulation is proposed in (Simon and Divsalar, 1992) to remove the CFO effect.

This article is structured as follows. In the first section, the overview of differential modulation and demodulation is described, followed by the sequence-based differential detection techniques that improve the performance of differential detection. In next section, the concept of double differential technique in the presence of carrier offset, omitting the channel knowledge is presented, and finally, conclusions and future directions are drawn.

**DIFFERENTIAL DETECTION TECHNIQUES**

This section presents an overview of the classical differential encoded detection techniques, MSDD as well as Multiple-Symbol Differential Sphere Detection (MSDSD) technique in the view of cooperative diversity.

As discussed, the differential encoding is one of the most efficient techniques that eliminates the need for channel estimation of all the transmission links in the networks. It is observed that although the differential modulation saves the bandwidth and the power wastage, the differential modulation and demodulation suffers from a performance loss as compared to the ideal (i.e. perfect channel estimation) coherent modulation system.

---
[1]Phase-shift keying (PSK) is a digital modulation method wherein the phase of a reference signal is varied, or modulated to convey information.

However, in practical time-varying scenario, it is a difficult task to attain the channel phase information perfectly in coherent detection due to the unknown CSI (Chen and Laneman, 2006). For that reason, the reliability (including but not limited to the power efficiency, throughput, network security, etc.) of the wireless communication may not be achieved when the channel is not estimated properly.

Figure 1 represents the block diagram of a single differential encoder and decoder (El-Hajjar et al., 2010).

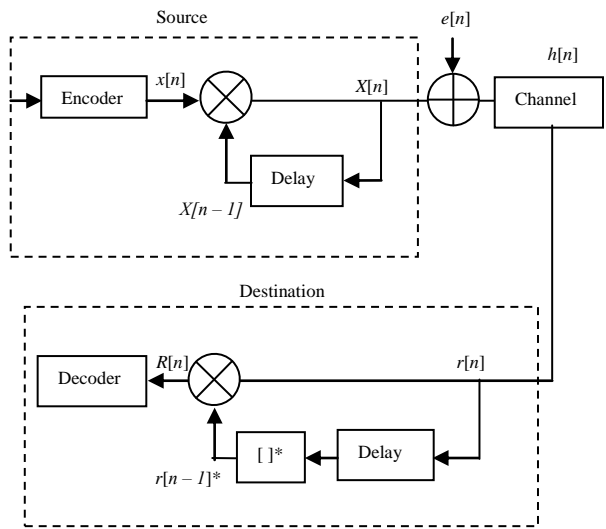

Fig. 1. A Block Diagram of a Differential Encoder and Decoder (El-Hajjar et al., 2010).

The data in the form of bits pattern is first mapped to symbols $x[n]$ when the DPSK is employed. Then the symbols are encoded differentially. At time instance $n$, the differentially modulated symbol $X[n]$ is transmitted as shown in figure 1. $X[n]$ can be obtained as $x[n] \times X[n-1]$, where $X[n-1]$ is the delayed version of the differentially encoded symbol at time instant $n-1$. $r[n]$ denotes the received signal at the destination. Assuming that $h[n]$ is the channel between the source and destination and $e[n]$ refers to the noise, $r[n]$ is acquired as $h[n]X[n] + e[n]$. The received symbols are detected by computing and finding the closest constellation of $r[n] \times r[n-1]^*$, where $[\ ]^*$ signifies the complex conjugate[2] operation. Instead of comparing the phase relative to a reference signal (i.e. coherent receiver), the decoder determines the phase change of the previous and currently detected signal. DPSK is beneficial as compare to Pulse Shift Keying (PSK), since the modulation schemes do not require exact phase reference for the received signal. But, it produces errors due to comparison between the two noisy signals at the destination in case that the channel amplitude and phase changes slowly over a period for two consecutive time slots (Tarokh and Jafarkhani, 2000; El-Hajjar et al., 2010). Despite of the expense for performance loss, differential detection is still preferable in wireless networks, particularly in cooperative diversity system when the channel estimation poses a challenge in the fading environment.

A conventional cooperative diversity system comprises of a source, a relay and a destination as depicted in Figure 2 (Zhao and Li, 2005). In this strategy, neighbouring node that acts as a relay helps to forward the source signal to an intended destination. The cooperative diversity considers two stage transmission, where signals between the source-relay, relay-destination and source-destination are transmitted in different time slots. In the first stage, the source broadcasts its signals to both relay and destination. Then, during the second stage, the received signals are processed using relay protocols such as Amplify-and-Forward (AF), Decode-and Forward (DecF) and Detect-and Forward (DetF). The relay then retransmits its differentially encoded signal towards the destination and the destination receives both the relayed and directly transmitted signals. The two replicas of the source signals are combined and decoded at the destination in order to achieve diversity gain[3] and to thus improve the Bit Error Rate (BER).

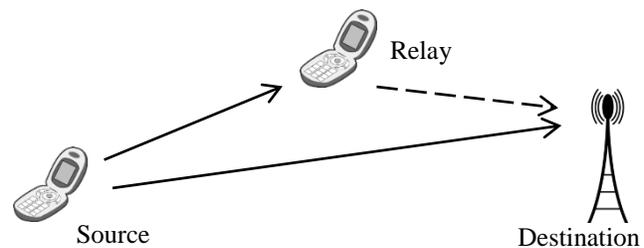

Fig. 2. Cooperative Communication for a Single Source, Relay and Destination (Zhao and Li, 2005).

In coherent communication, perfect channel estimation is required at both the relay and destination in the first stage of transmission. Additionally, during the second stage, the destination has to estimate the CSI for the relay-destination channel, which incurs loss in the available bandwidth due to the transmission of pilot symbols. In order to circumvent the channel estimation, differential modulation based schemes, namely Differential Amplify-and Forward (DAF) and Differential Decode-and Forward (DDecF) for cooperative systems were proposed in (Zhao and Li, 2007). The information bits at the source are differentially encoded using Binary Phase Shift Keying (BPSK) constellation and broadcasted to the relay and directly to the destination. The relay then amplifies or decodes the received signal and retransmits the signal towards the destination. At the destination, the detection and combining processes follow a Maximum Likelihood (ML) approach for DAF and Piecewise-Linear (PL) approach for DDecF. The data received from both the relay and source is decoded with the differential linear combiner employing the average Signal-to-Noise Ratio (SNR) instead of instantaneous SNR without having explicit reliance on the unknown channels. PL is proposed as an accurate approximation of the nonlinear ML detectors for DDecF protocol network. A framework associated with analytical out comes in terms of average BER, outage probability and Probability Density Function (PDF) for the DAF and DDecF schemes have been developed and outperformed the conventional non-cooperative network.

---

[2]The complex conjugate number constitutes of real and imaginary part wherein the real part equals for a given complex number and the imaginary part corresponds in magnitude but opposite in sign.

[3]The diversity gain is the performance in link reliability due to diversity scheme whereby multiple replicas of information signal is received via independent fading links.

A differential scheme designed for two-user cooperative communication system based on differential space-time block codes was proposed in (Tarasak *et al*., 2005). The cooperative system comprises of two cooperative users that acts as source as well as relay at the same time and a destination. As depicted in Figure 3, the transmission is divided into three time slots. In the first time slot, User 1 transmits its differentially encoded data to User 2 and at the same time directly to the destination via independent paths.

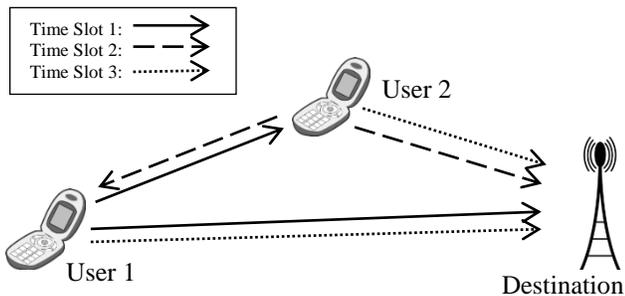

Fig. 3. Two-user Cooperative Diversity System (Poramate Tarasak *et al*., 2005).

During the second time slot, User 2 performs differential encoding process to the data which is transmitted to User 1 and the destination. In the third time slot, both users act as relays to differentially decode as well as re-encode other user's messages. At the same time, the negative and conjugate operation[4] is performed before transmitting the relayed signals towards the destination with the structure that is similar to the Alamouti's scheme (Alamouti, 1998). According to Alamouti's matrix scheme,the directly transmitted symbols and relayed symbols are constituted in matrix form wherein the components of the first row matrix represent the directly transmitted symbols whilst the second row matrix components represent the relayed symbols. Since the CSI is unknown at the destination,

---

[4] The first user transmits the negative and conjugate relayed signal while the second user transmits the conjugated signal based on Alamouti's scheme.

DecF protocol is employed and CRC bits are assumed to be added to both users before the encoding process. Upon receiving the encoded source signal, each user that act as relay performs CRC check after differential decoding and retransmits the correct re-encoded signal towards the destination. The received signals at the destination are differentially decoded with the addition of noise, symbol error, and delay. The scheme achieves second-order diversity efficiently and has an acceptable performance if the noise is small.

The existing differential cooperative schemes can are further extended to multi-node networks such as, in the application of wireless sensor and ad hoc networks. In (Himsoon *et al*., 2008), differential schemes with AF and DecF cooperative networks have been investigated for multi cooperative relays cooperation systems. For the DAF, each relay simply amplifies the obtained signal from the source and transmits the amplified signal to the intended destination with the assumption that the channel variances are available at the destination. The combined signals are differentially decoded by referring to the detection rule in (Ricklin and Zeidler, 2009). Whereas, for the DDecF scheme, each relay decodes the received signal using the detection rule in (Ricklin and Zeidler, 2009) and only retransmit the correctly decoded symbol towards the desired destination. At the destination, a threshold-based decision is applied to determine the highly accurate information having signal from each one of the relay link to be combined for joint decoding with the directly transmitted signal from the source. It is observed that the simulated BER matches the exact theoretical BER benchmark and achieve diversity orders as number of relays $M$ increases. However, a performance gap is observed between the DDAF and DDecF schemes and the coherent transmission scheme. Generally, the technique performance degrades due to its inability to recover the source information especially in fast fading environment.

Therefore, the MSDD based on ML criterion has been proposed in (Divsalar and Simon, 1990) to fill the performance gap between the coherent detection assuming perfect knowledge estimation at the destination and the differential scheme. This technique can be carried out simply by increasing the amount of symbols to form a block of $N$ consecutive PSK symbols for joint detection of $N-1$ consecutively transmitted data symbols process, rather than symbol-based detection scheme. It is observed that the performance of the error-rate significantly improves with the increase of the observation window size $N$. However, as the number of symbols grow, the detection complexity also increases exponentially. Several suboptimal algorithms were proposed in (Xiaofu and Songgeng, 1998; Tarasak and Bhargava, 2002; Li, 2003) to reduce the detection complexity of the MSDD based on subset search concept. First, a block sequence as a reference is used to find the maximum candidate list with two consecutive symbols. Then, these candidates which maximize the MSDD metric are searched and tested with increased window size. However, the suboptimal detectors perform only the reduction of the complexity analysis without promising for the detection decision.

On the contrary, a generalized differential modulation for AF wireless relay networks was suggested in (Fang *et al.*, 2009) to combat the complexity of MSDD technique. In the proposed scheme, the bit sequences are divided into multiple blocks, wherein the first symbol of the block is differentially encoded. The other symbols of the block are differentially encoded in line with the initial symbol of the similar block. At the destination, combined signals from the source and relays are decoded differentially. It is observed that the performance of the system improves as the length of the block increases over slow Rayleigh fading channels, given the channels remain unchanged for a block period. Another simpler alternative of MSDD with the application of Sphere Decoding (SD) was proposed in (Lampe *et al.*, 2005). The concept was popularly modelled in MIMO system, Code Division Multiple Access (CDMA) network and also advocated in the cooperative communication system. Frame-based cooperation is carried out rather than symbol-based due to the performance degradation as the environment becomes more time-selective (Ricklin and Zeidler, 2009). Firstly, in MSDSD, a decision metric based on the ML decoding MSDD is obtained by applying the Cholesky factorization[5] of the channel statistics.

Secondly, the decision metric is then adapted with the SD concept. Based on SD approach, the vector search for the ML estimation is limited to a spherical region within certain radius R, cantered at the received vector.

Thirdly, the radius R is dynamically updated when the candidate vector is found applying the Schnorr-Euchner search strategy[6] (Schnorr and Euchner, 1994) and SD is repeated with new radius R. Finally, the MSDSD process is done when no vector is found within the sphere of radius R. MSDSD has been extensively extended in many works owing to the extension decision of the window length. MSDSD for the DAF cooperative system has been proposed in (Wang and Hanzo, 2009). The studies have shown that the MSDD and MSDSD demonstrate significant improvement as compared to the conventional differential schemes for data recovery detection.

Most of the works found in the previous literature devise differential transmission schemes that bypass channel estimation at the destination assuming perfect synchronization. However, for cooperative diversity where the source, relay and destination may be in mobility state, synchronization will never be perfectly matched due to the CFO effect. Therefore, double-differential modulation is proposed considering CFO effect without the knowledge of CSI.

**DOUBLE-DIFFERENTIAL TECHNIQUE**

In practical environment, with the presence of CFO, the differential detection may fail to achieve the desired performance particularly in a wireless mobile network that causes the Doppler effect. In addition, due to the mobility of the nodes, the oscillators of the transmitter(s), relay(s) and the source(s) in cooperative diversity are experiencing difficulties to match perfectly (El-Hajjar *et al.*, 2010). Thus, double-differential (DD) technique has been proposed in (Bhatnagar *et al.*, 2008) to mitigate the problem of attaining good performance in the absence of channel knowledge. The DD block diagram is depicted in Figure 4.

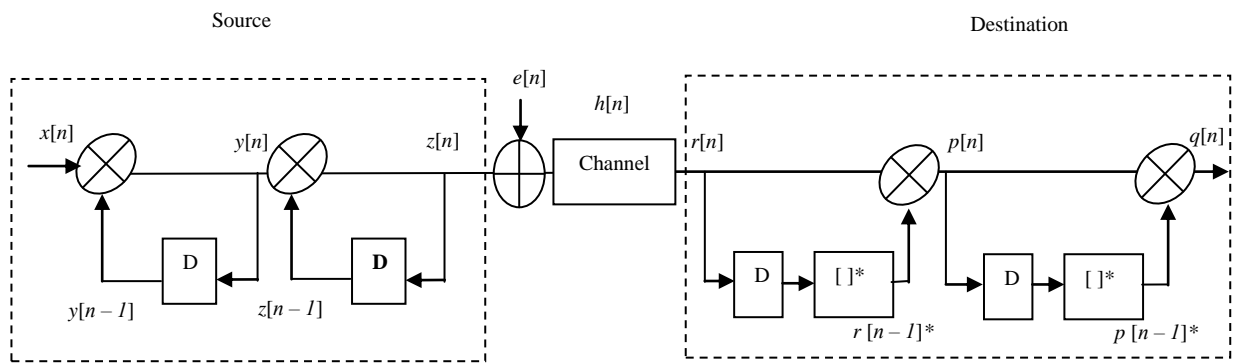

Fig. 4. A Block Diagram of Double Differential Encoder and Decoder (Bhatnagar *et al.*, 2008).

---

[5]The Cholesky factorization is a method to solve numerical problems by factoring a matrix into a product of a lower triangular matrix with its conjugate transpose.

[6]The Schnorr-Euchner enumeration calculates the connecting branch weights of a tree-based search to locate a point that is close to the target. Once the point is discovered, the search radius is adaptively reduced which lower the complexity level of the search.

A block diagram of DD encoder and decoder is depicted in Figure 4 wherein D denotes delay and $x[n]$ represents the encoded signal at time instance $n$. The transmitted signal at time instance $n$ can be attained as $z[n] = x[n] \times y[n-1] \times z[n-1]$, where, the classic single differential modulation is added with an additional delay stage $z[n-1]$. Considering a point-to-point transmission with carrier offset, the received signal is described by $r[n] = he^{wn}r[n] + e[n]$, where, $h$ is the channel gain and $w$ is the unknown carrier offset with $e[n]$ as the AWGN noise. $q[n]$ denotes the decision variable and is represented by $p[n]p[n-1]^*$, where $p[n] = r[n]r[n-1]^*$. Although cooperative communication systems achieve diversity gain, the systems are susceptible to CFO. Stoica *et al.* (2003) has derived the double differential demodulation based on the simple heuristic detector preferred over the complex Maximum-Likelihood Detection (MLD). The results signify that the heuristic detector matches with the MLD. The double differential modulation approach is further investigated in (Bhatnagar *et al.*, 2008) for the cooperative diversity employing AF protocol in the presence of CFO.

An Emulated Maximum Ratio Combining (EMRC) scheme exploiting the channel variances is proposed at the destination to decode the combine received signals in the absence of channel knowledge. The analysis of the BER was provided which outperformed the coherent cooperative system.

## CONCLUSION

Generally, the idea of allowing neighbouring nodes helping each other by cooperating among themselves is an alternative way to increase data transmission rate and to enhance the system reliability in wireless communication networks. This article gives an overview of differential transmission techniques mainly when the CSI is unknown at the destination. The differential encoding and decoding concept is based on the difference between the phases of two consecutive received symbols that omit channel estimation. It is shown that even without the knowledge of the CSI at the receiver, the detection techniques guarantee good performances in terms of system reliability in flat-fading environment. In contrast, the coherent receivers require the perfect and exact channel knowledge. The CSI can be obtained by the transmission of extra pilot symbols, resulting in the wastage of the transmission power as well as the available transmission bandwidth. Hence, the differential schemes based on symbol-by-symbol and multiple symbol transmission are preferable particularly in a time-varying environment or when the channel is inaccurately estimated, which deteriorates the performance of the coherent systems. In the presence of CFO, double differential approach was investigated to provide reliable communication link where the differential receivers fails to perform. The overview of the comparison among different types of detection techniques highlight the interesting issues pertaining to foreseeable future study such as detection techniques to eliminate CFO without channel knowledge for cooperative diversity system.

## REFERENCES

Alamouti, SM. 1998. A simple transmit diversity technique for wireless communications. IEEE Journal on Selected Areas in Communications.16(8):1451-1458.

Annavajjala, R., Cosman, PC. and Milstein, LB. 2005. On the performance of optimum noncoherent amplify-and-forward reception for cooperative diversity. Proceedings - IEEE Military Communications Conference MILCOM. 1-9.

Bhatnagar, MR., Hjrongnes, A. and Song, L. 2008. Cooperative communications over flat fading channels with carrier offsets: A double-differential modulation approach. Eurasip Journal on Advances in Signal Processing. (531786):1-11.

Chen, D. and Laneman, JN. 2006. Modulation and demodulation for cooperative diversity in wireless systems. IEEE Trans. Wireless Commun. 5(7):1785-1794.

Divsalar, D. and Simon, MK. 1990. Multiple-symbol differential detection of MPSK. IEEE Transactions on Communications. 38(3):300-308.

El-Hajjar, M., Hanzo, L., Versus, C. and Communication, N. 2010. Dispensing with channel estimation. IEEE Vehicular Technology Magazine. 42-48.

Fang, Z., Li, L., Bao, X. and Wang, Z. 2009. Generalized differential modulation for amplify-and-forward wireless relay networks. IEEE Transactions on Vehicular Technology. 58(6):3058-3062.

Himsoon, T., Siriwongpairat, WP., Su, W. and Liu, KJR. 2008. Differential modulations for multinode cooperative communications. IEEE Transactions on Signal Processing. 56(7 I):2941-2956.

Lampe, L., Schober, R., Pauli, V. and Windpassinger, C. 2005. Multiple-symbol differential sphere decoding. IEEE Transactions on Communications. 53(12):1981-1985.

Laneman, JN., Tse, DNC. and Wornell, GW. 2004. Cooperative diversity in wireless networks: Efficient protocols and outage behavior. IEEE Transactions on Information Theory. 50(12):3062-3080.

Laneman, JN. and Wornell, GW. 2003. Distributed space–time-coded protocols for exploiting cooperative diversity in wireless networks. IEEE Transactions on Information Theory. 49(10):2415-2425.

Li, B. 2003. A new reduced-complexity algorithm for multiple-symbol differential detection. IEEE Commu. 7(6):269-271.

Nosratinia, A., Hunter, TE. and Hedayat, A. 2004. Cooperative communication in wireless networks. Communications Magazine. IEEE. 42(10):74-80.

Proakis, JG. 2001. Digital Communications (4$^{th}$ edi.). McGraw-Hill, New York, USA.

Ricklin, N. and Zeidler, JR. 2009. Block detection of multiple symbol DPSK in a statistically unknown time-varying channel. IEEE International Conference on Communications. 1:0-4.

Schnorr, CP. and Euchner, M. 1994. Lattice basis reduction: Improved practical algorithms and solving subset sum problems. Mathematical Programming. 66(1-3):181-199.

Sendonaris, A., Erkip, E. and Aazhang, B. 2003. User cooperation diversity. IEEE Transactions on Communications. 51(11):1-9.

Simon, MK. and Divsalar, D. 1992. On the implementation and performance of single and double differential detection schemes. IEEE Transactions on Communications. 40(2):278-291.

Skylar, B. 1988. Digital Communications Fundamentals and Applications. Prentice-Hall International Edition. New Jersey, USA.

Stoica, P., Liu, JLJ. and Li, JLJ. 2003. Maximum likelihood double differential detection clarified. 4$^{th}$ IEEE Workshop on Signal Processing Advances in Wireless Communications - SPAWC. (9):393-397.

Tarasak, P. and Bhargava, VK. 2002. Reduced complexity multiple symbol differential detection of space-time block code. IEEE Wireless Communications and Networking Conference Record. WCNC. 21(C):505-509.

Tarasak, P., Minn, H. and Bhargava, VK. 2005. Differential modulation for two-user cooperative diversity systems. IEEE Journal on Selected Areas in Communications. 23(9):1891-1900.

Tarokh, V. and Jafarkhani, H. 2000. A differential detection scheme for transmit diversity. IEEE Journal on Selected Areas in Communications. 18(7):1169-1174.

Tian, J., Zhang, Q. and Yu, F. 2011. Non-coherent detection for two-way AF cooperative communications in fast rayleigh fading channels. IEEE Transactions on Communications. 59(10):2753-2762.

Wang, L. and Hanzo, L. 2009. The amplify-and-forward cooperative uplink using multiple-symbol differential sphere-detection. IEEE Signal Processing Letters. 16(10):913-916.

Xiaofu, W. and Songgeng, S. 1998. Low complexity multi symbol differential detection of MDPSK over flat correlated rayleigh fading channels. Electronics Letters. 34(21):2008-2009.

Zhao, Q. and Li, H. 2005. Performance of differential modulation with wireless relays in Rayleigh fading channels. IEEE Communications Letters. 9(4):343-345.

Zhao, Q. and Li, H. 2007. Differential modulation for cooperative wireless systems. IEEE Transactions on Signal Processing. 55(5):2273-2283.

Zhu, Y., Kam, PY. and Xin, Y. 2007. Non-coherent detection for amplify-and-forward relay systems in a rayleigh fading environment. IEEE GLOBECOM, IEEE Global Telecommunications Conference. 1658-1662.